# 18-00590
# REAL-TIME CRASH RISK ANALYSIS OF URBAN ARTERIALS INCORPORATING BLUETOOTH, WEATHER, AND ADAPTIVE SIGNAL CONTROL DATA

**Jinghui Yuan (Corresponding Author), M.S.,** PhD student
Department of Civil, Environmental & Construction Engineering, University of Central Florida
Orlando, FL 32816
Email: jinghuiyuan@knights.ucf.edu

**Mohamed Abdel-Aty, Ph.D., P.E.**
Pegasus Professor and Chair
Department of Civil, Environmental & Construction Engineering, University of Central Florida
Orlando, FL 32816
Phone: (407) 823-4535 Fax: (407) 823-3315
Email: M.Aty@ucf.edu

**Ling Wang, Ph.D.**
[1]Department of Civil, Environmental and Construction Engineering, University of Central Florida，
Orlando, FL 32816，USA
[2]College of transportation engineering, Tongji University,
1239 Siping Rd, Yangpu Qu, Shanghai Shi, 200000, China
E-mail: LingWang@knights.ucf.edu

**Jaeyoung Lee, Ph.D.**
Department of Civil, Environmental & Construction Engineering, University of Central Florida
Orlando, FL 32816
Email: jaeyoung@knights.ucf.edu

**Xuesong Wang, PhD**
School of Transportation Engineering, Tongji University
4800 Cao'an Road, Jiading District, Shanghai, 201804, China
Email: wangxs@tongji.edu.cn

**Rongjie Yu, PhD**
School of Transportation Engineering, Tongji University
4800 Cao'an Road, Jiading District, Shanghai, 201804, China
Email: yurongjie@tongji.edu.cn

**INTRODUCTION**

Urban arterials play a critical role in the road network system as they provide the high-capacity network for travel within urban areas as well as the access to roadside activities. Meanwhile, urban arterials suffer from serious traffic safety issues. Take Florida as an example, over 51% of crashes have occurred on urban arterials in 2014. Substantial efforts have been made by the previous researchers to reveal the relationship between crash frequency on urban arterials and all the possible contributing factors such as roadway geometric, and traffic characteristics, etc. (*1-4*). However, these studies were conducted based on static and highly aggregated data (e.g., Annual Average Daily Traffic (AADT), annual crash frequency).

Recently, an increasing number of studies investigated the crash likelihood on freeways by using real-time traffic and weather data (*5-14*). However, little research has been conducted on the real-time safety analysis of urban arterials (*15; 16*). This may be due to the substantial difference in the traffic flow characteristics between urban arterials and freeways. More specifically, the interrupted traffic flow on urban arterials is highly controlled by the traffic signals (*17; 18*), which is quite different from the free flow on freeways. Therefore, the crash risk on urban arterials might be associated with not only real-time traffic flow characteristics but also the real-time signal phasing, which has not been considered in the previous research (*15; 16*). Moreover, those pioneering studies on the real-time safety analysis of urban arterials were based on one-hour aggregated traffic parameters prior to crash occurrence, which is not really exact "real-time" as the traffic flow are likely to differ within one hour.

Above all, this study aims to investigate the relationship between crash occurrence on urban arterials and real-time traffic, signal phasing, and weather characteristics by utilizing data from multiple sources, i.e., Bluetooth, weather, and adaptive signal control datasets.

**DATA PREPARATION**

A total of four datasets were used: (1) 113 crashes from March, 2017 to June, 2017 provided by Signal Four Analytics (S4A); (2) travel speed data collected by 23 IterisVelocity Bluetooth detectors; (3) signal phasing and 15-minute interval traffic volume provided by 23 adaptive signal controllers; (4) weather characteristics collected by the nearest airport weather station. All the real-time traffic data were extracted for a period of 20 minutes (divided into four 5-minute time slices) before the time of crash occurrence. As shown in Figure 1, the composition of traffic volume for the upstream and downstream intersections are different. Three weather related variables (rainy weather indicator, visibility, and hourly precipitation) were collected from the nearest airport weather station, which is located at the Orlando international airport.

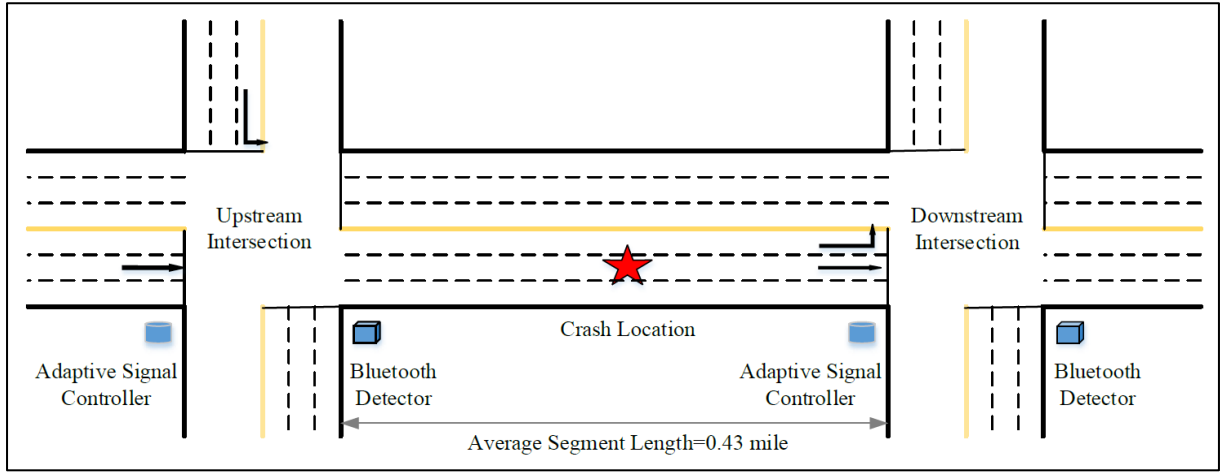

**FIGURE 1 Arrangement of Bluetooth Detectors and Adaptive Signal Controller**

Matched case-control design was adopted in this study to create a non-crash dataset. Different control-to-case ratio from 1:1 to 10:1 were examined, and 8:1 was found to have the best model performance. Consequently, data for 8 non-crash cases for the same road segment, time of day, and day of week were also extracted. The final dataset included 113 crashes and 904 non-crash cases.

## METHODOLOGY
### Bayesian Conditional Logistic Model

Suppose that there are $N$ strata with 1 crash ($y_{ij}$=1) and $m$ non-crash cases ($y_{ij}$=0) in stratum $i$, $i$=1, 2, …, $N$. Let $p_{ij}$ be the probability that the $j$th observation in the $i$th stratum is a crash; $j$=0, 1, 2, …, $m$. This crash probability could be expressed as:

$$y_{ij} \sim Bernoulli(p_{ij}) \qquad (1)$$

$$logit(p_{ij}) = \alpha_i + \beta_1 X_{1ij} + \beta_2 X_{2ij} + \cdots + \beta_k X_{kij} \qquad (2)$$

Where $\alpha_i$ is the intercept term for the $i$th stratum; $\boldsymbol{\beta} = (\beta_1, \beta_2, \dots, \beta_k)$ is the vector of regression coefficients for $k$ independent variables, and all the $\boldsymbol{\beta}$ coefficients are set up with non-informative priors as following normal distributions Normal (0, 1000); $\boldsymbol{X_{ij}} = (X_{1ij}, X_{2ij}, \dots, X_{kij})$ is the vector of $k$ independent variables.

In order to take the stratification in the analysis of the observed data, the stratum-specific intercept $\alpha_i$ is considered to be nuisance parameters, and the conditional likelihood for the $i$th stratum would be expressed as (*19*):

$$l_i(\boldsymbol{\beta}) = \frac{\exp(\sum_{u=1}^{k} \beta_u X_{ui0})}{\sum_{j=0}^{m} \exp(\sum_{u=1}^{k} \beta_u X_{uij})} \qquad (3)$$

And the full conditional likelihood is the product of the $l_i(\beta)$ over $N$ strata,

$$L(\boldsymbol{\beta}) = \prod_{i=1}^{N} l_i(\boldsymbol{\beta}) \qquad (4)$$

Since the full conditional likelihood is independent of stratum-specific intercept $\alpha_i$, thus Equation 2 cannot be used to estimate the crash probabilities. However, the $\boldsymbol{\beta}$ coefficients can be estimated by Equation 4. These estimates are the log-odd ratios of corresponding variables and can be used to approximate the relative risk of a crash. Furthermore, the log-odds ratios can also be used to develop a prediction model under this matched case-control analysis.

Yuan *et al.* 4

Suppose two observation vectors $\boldsymbol{X_{i1}} = (X_{1i1}, X_{2i1}, \ldots, X_{ki1})$ and $\boldsymbol{X_{i2}} = (X_{1i2}, X_{2i2}, \ldots, X_{ki2})$ from the *i*th strata, the odds ratio of crash occurrence caused by observation vector $\boldsymbol{X_{i1}}$ relative to observation vector $\boldsymbol{X_{i2}}$ could be calculated as:

$$\frac{p_{i1}/(1-p_{i1})}{p_{i2}/(1-p_{i2})} = \exp[\sum_{u=1}^{k} \beta_u (X_{ui1} - X_{ui2})] \tag{5}$$

The right hand side of Equation 5 is independent of $\alpha_i$ and can be calculated using the estimated $\boldsymbol{\beta}$ coefficients. Thus, the above relative odds ratio may be utilized for predicting crash occurrences by replacing $\boldsymbol{X_{i2}}$ with the vector of the independent variables in the *i*th stratum of non-crash cases. Let $\overline{\boldsymbol{X}}_i = (\overline{X}_{1i}, \overline{X}_{2i}, \ldots, \overline{X}_{ki})$ denote the vector of mean values of non-crash cases of the *k* variables within the *i*th stratum. Then the odds ratio of a crash relative to the non-crash cases in the *i*th stratum may be approximated by:

$$\frac{p_{i1}/(1-p_{i1})}{p_{\bar{i}}/(1-p_{\bar{i}})} = \exp[\sum_{u=1}^{k} \beta_u (X_{ui1} - \overline{X}_{ui})] \tag{6}$$

**Bayesian Inference and Model Comparisons**
Full Bayesian inference was employed in this study. The Deviance Information Criterion (DIC) can be used to compare complex models by offering a Bayesian measure of model fitting and complexity (*20*). DIC is defined as outlined in Equation 12:

$$DIC = \overline{D(\theta)} + p_D \tag{11}$$

Where $D(\theta)$ is the Bayesian deviance of the estimated parameter, and $\overline{D(\theta)}$ is the posterior mean of $D(\theta)$. $\overline{D(\theta)}$ can be viewed as a measure of model fit, while $p_D$ denotes the effective number of parameters and indicates the complexity of the models.

For the model goodness-of-fit, AUC, which is area under Receiver Operating Characteristic (ROC) curve was also adopted. It is worth noting that the classification result of Bayesian conditional logistic model are based on the predicted odds ratio, which may be larger than 1. In order to be comparable with other models, all the predicted odds ratios were divided by the maximum odds ratio to create normalized odds ratios. Later on, the normalized odds ratios were used to create the classification result based on different threshold from 0 to 1.

**MODELING RESULTS**
Since all the traffic parameters during time slices 3 and 4 were insignificant, and the rainy weather indicators are identical in the four datasets, thus only three models were presented in Table 2.

**TABLE 1 Model Results of Bayesian Conditional Logistic Regression Models based on Different Time Slices**

| Parameter | Time Slice 1 (0-5 minute) | | Time Slice 2 (5-10 minute) | | Time Slice 3 (10-15 minute) | |
|---|---|---|---|---|---|---|
| | Mean (95% BCI) | Hazard Ratio | Mean (95% BCI) | Hazard Ratio | Mean (95% BCI) | Hazard Ratio |
| Avg_speed | **-0.056 (-0.084, -0.029)** | 0.946 | **-0.028 (-0.058, 0.004)*** | 0.972 | - | - |
| Up_Vol | - | | **0.008 (-0.001, 0.017)*** | 1.008 | - | - |
| Rainy | **0.799 (-0.048, 1.582)*** | 2.223 | **0.751 (-0.015, 1.519)*** | 2.119 | **0.815 (0.041, 1.551)** | 2.258 |
| AUC | 0.607 | | 0.572 | | 0.515 | |

*Note: Mean (95% BCI) values marked in bold are significant at the 0.05 level; Mean (95% BCI) values marked in bold and noted by * are significant at the 0.1 level.*

As presented in Table 1, the model comparison results based on AUC values indicate that the slice 1 model (0-5 minute interval) performs the best, followed by the slice 2 (5-10 minute interval) model. On the other hand, slice 2 model performs the best in terms of the number of significant variables. Finally, the slice 2 model was selected to conduct further interpretation and model comparison.

Based on the estimation results of the slice 2 model, three variables were found to be significantly associated with the crash occurrence on urban arterials: (1) the negative coefficient (-0.028) of average speed indicates that higher average speed tends to decrease the crash risk, which is consistent with other studies (*6; 7; 21-25*). The odds ratio of 0.972 means that one-unit increase in the average speed would decrease the odds of crash occurrence by 2.8%; (2) upstream volume was found to be positively correlated with crash likelihood, and the odd ratio of 1.008 indicates that one-unit increase in upstream volume would lead to an increase of 0.8% in the odds of crash occurrence; (3) rainy weather indicator also has a positive coefficient, the odd ratio of 2.119 means that odds of crash occurrence under rainy condition is 111.9% higher than normal conditions, which is in line with previous studies (*7; 10*).

Furthermore, both Bayesian logistic model and Bayesian random effect logistic model were developed based on time slice 2 dataset, the model comparison results are as shown in Table 2. Based on the DIC and AUC values, it is obvious to conclude that the Bayesian conditional logistic model performs much better than the other two models.

**TABLE 2 Model Comparison Results based on Time Slice 2**

| Parameter | Bayesian conditional logistic model | | Bayesian logistic model | | Bayesian random effect logistic model | |
|---|---|---|---|---|---|---|
| | Mean (95% BCI) | Hazard Ratio | Mean (95% BCI) | Hazard Ratio | Mean (95% BCI) | Hazard Ratio |
| Intercept | - | - | **-2.021 (-2.717, -1.375)** | - | **-2.044 (-2.728, -1.342)** | - |
| Avg_speed | **-0.028 (-0.058, 0.004)*** | 0.972 | -0.011 (-0.030, 0.007) | 0.989 | -0.011 (-0.030, 0.008) | 0.989 |
| Up_Vol | **0.008 (-0.001, 0.017)*** | 1.008 | 0.002 (-0.002, 0.006) | 1.002 | 0.002 (-0.002, 0.006) | 1.002 |
| Rainy | **0.751 (-0.015, 1.519)*** | 2.119 | **0.667 (-0.066, 1.397)*** | 1.949 | **0.648 (-0.067, 1.387)*** | 1.912 |
| Random effect (tau) | - | - | - | - | 393.5 | - |
| DIC | 491.418 | | 711.433 | | 713.239 | |
| AUC | 0.572 | | 0.562 | | 0.562 | |

*Note: Mean (95% BCI) values marked in bold are significant at the 0.05 level; Mean (95% BCI) values marked in bold and noted by * are significant at the 0.1 level.*

## CONCLUSION AND DISCUSSION

The results of the slice 2 model indicate that the average speed, upstream volume, and rainy weather indicator are significantly associated with the crash risk on urban arterials. In general, these finding are consistent with previous studies, in which the average speed was found to have significant negative impact on crash occurrence (*6; 7; 21-25*), while upstream volume (*23; 25; 26*) and adverse weather (*7; 10*) were found to be positively correlated with crash risk. Surprisingly, the coefficient of variation in speed is insignificant, this could be explained in that the average number of vehicles detected by the Bluetooth detector within 5-minute interval is about 6, which might be too small to capture the variation in speed.

Compared with the previous study on the real-time safety analysis of urban arterials (*27*), which found that the 1 hour variation in both occupancy and volume were significantly associated with crash likelihood, which is quite different from our study. This might be because the 1 hour aggregated traffic parameters can hardly represent the actual short-term traffic status such as speed and volume prior to crash occurrence, while it can capture the variation in traffic flow. This comparison implies that the traffic parameters should be aggregated based on more appropriate time interval, which can not only represent the short-term traffic status but also capture the variation in traffic flow.

Furthermore, the model comparison results indicate that the Bayesian conditional logistic model performs much better than the other two models, which means Bayesian conditional logistic model is more preferable in the context of the matched case-control dataset. The AUC value of 0.572 implies that the model is still hard to be applied to the arterial traffic management system, however, the estimation results provide profound insights for traffic engineers to understand the relationship between crash risk and real-time traffic characteristics and weather conditions. In future studies, more advanced machine learning techniques could be applied to improve the predictive performance.

**REFERENCES**

1. El-Basyouny, K., and T. Sayed. Accident prediction models with random corridor parameters. *Accident Analysis & Prevention,* Vol. 41, No. 5, 2009, pp. 1118-1123.
2. Gomes, S. V. The influence of the infrastructure characteristics in urban road accidents occurrence. *Accident Analysis & Prevention,* Vol. 60, 2013, pp. 289-297.
3. Greibe, P. Accident prediction models for urban roads. *Accident Analysis & Prevention,* Vol. 35, No. 2, 2003, pp. 273-285.
4. Wang, X., T. Fan, M. Chen, B. Deng, B. Wu, and P. Tremont. Safety modeling of urban arterials in Shanghai, China. *Accident Analysis & Prevention,* Vol. 83, 2015, pp. 57-66.
5. Abdel-Aty, M., N. Uddin, A. Pande, F. Abdalla, and L. Hsia. Predicting freeway crashes from loop detector data by matched case-control logistic regression. *Transportation Research Record: Journal of the Transportation Research Board*, No. 1897, 2004, pp. 88-95.
6. Abdel-Aty, M. A., H. M. Hassan, M. Ahmed, and A. S. Al-Ghamdi. Real-time prediction of visibility related crashes. *Transportation Research Part C: Emerging Technologies,* Vol. 24, 2012, pp. 288-298.
7. Ahmed, M., M. Abdel-Aty, and R. Yu. Assessment of Interaction of Crash Occurrence, Mountainous Freeway Geometry, Real-Time Weather, and Traffic Data. *Transportation Research Record: Journal of the Transportation Research Board,* Vol. 2280, 2012, pp. 51-59.
8. Lee, C., B. Hellinga, and F. Saccomanno. Real-time crash prediction model for application to crash prevention in freeway traffic. *Transportation Research Record: Journal of the Transportation Research Board*, No. 1840, 2003, pp. 67-77.
9. Oh, C., J.-S. Oh, S. Ritchie, and M. Chang. Real-time estimation of freeway accident likelihood. Presented at 80th Annual Meeting of the Transportation Research Board, Washington, DC, Washington, D.C., 2001.
10. Xu, C., A. P. Tarko, W. Wang, and P. Liu. Predicting crash likelihood and severity on freeways with real-time loop detector data. *Accident Analysis & Prevention,* Vol. 57, 2013, pp. 30-39.
11. Xu, C., W. Wang, and P. Liu. Identifying crash-prone traffic conditions under different weather on freeways. *Journal of Safety Research,* Vol. 46, 2013, pp. 135-144.
12. Yu, R., and M. Abdel-Aty. Analyzing crash injury severity for a mountainous freeway incorporating real-time traffic and weather data. *Safety Science,* Vol. 63, 2014, pp. 50-56.
13. Yu, R., M. A. Abdel-Aty, M. M. Ahmed, and X. Wang. Utilizing microscopic traffic and weather data to analyze real-time crash patterns in the context of active traffic management. *IEEE transactions on intelligent transportation systems,* Vol. 15, No. 1, 2014, pp. 205-213.

14. Zheng, Z., S. Ahn, and C. M. Monsere. Impact of traffic oscillations on freeway crash occurrences. *Accident Analysis & Prevention,* Vol. 42, No. 2, 2010, pp. 626-636.
15. Theofilatos, A., and G. Yannis. A review of the effect of traffic and weather characteristics on road safety. *Accident Analysis & Prevention,* Vol. 72, 2014, pp. 244-256.
16. Theofilatos, A., G. Yannis, E. I. Vlahogianni, and J. C. Golias. Modeling the effect of traffic regimes on safety of urban arterials: The case study of Athens. *Journal of Traffic and Transportation Engineering (English Edition),* Vol. 4, No. 3, 2017, pp. 240-251.
17. Cai, Q., Z. Wang, L. Zheng, B. Wu, and Y. Wang. Shock wave approach for estimating queue length at signalized intersections by fusing data from point and mobile sensors. *Transportation Research Record: Journal of the Transportation Research Board*, No. 2422, 2014, pp. 79-87.
18. Wang, Z., Q. Cai, B. Wu, L. Zheng, and Y. Wang. Shockwave-based queue estimation approach for undersaturated and oversaturated signalized intersections using multi-source detection data. *Journal of Intelligent Transportation Systems,* Vol. 21, No. 3, 2017, pp. 167-178.
19. Hosmer Jr, D. W., S. Lemeshow, and R. X. Sturdivant. *Applied logistic regression*. John Wiley & Sons, Hoboken, New Jersey, 2013.
20. Spiegelhalter, D. J., N. G. Best, B. P. Carlin, and A. Van Der Linde. Bayesian measures of model complexity and fit. *Journal of the Royal Statistical Society: Series B (Statistical Methodology),* Vol. 64, No. 4, 2002, pp. 583-639.
21. Ahmed, M., M. Abdel-Aty, and R. Yu. Bayesian Updating Approach for Real-Time Safety Evaluation with Automatic Vehicle Identification Data. *Transportation Research Record: Journal of the Transportation Research Board,* Vol. 2280, 2012, pp. 60-67.
22. Ahmed, M. M., and M. A. Abdel-Aty. The Viability of Using Automatic Vehicle Identification Data for Real-Time Crash Prediction. *IEEE transactions on intelligent transportation systems,* Vol. 13, No. 2, 2012, pp. 459-468.
23. Shi, Q., and M. Abdel-Aty. Big Data applications in real-time traffic operation and safety monitoring and improvement on urban expressways. *Transportation Research Part C: Emerging Technologies,* Vol. 58, 2015, pp. 380-394.
24. Xu, C., P. Liu, W. Wang, and Z. Li. Evaluation of the impacts of traffic states on crash risks on freeways. *Accident Analysis & Prevention,* Vol. 47, 2012, pp. 162-171.
25. Yu, R., X. Wang, K. Yang, and M. Abdel-Aty. Crash risk analysis for Shanghai urban expressways: A Bayesian semi-parametric modeling approach. *Accident Analysis & Prevention,* Vol. 95, No. Pt B, 2016, pp. 495-502.
26. Yu, R., X. Wang, and M. Abdel-Aty. A Hybrid Latent Class Analysis Modeling Approach to Analyze Urban Expressway Crash Risk. *Accident Analysis & Prevention,* Vol. 101, 2017, pp. 37-43.
27. Theofilatos, A. Incorporating real-time traffic and weather data to explore road accident likelihood and severity in urban arterials. *Journal of Safety Research,* Vol. 61, 2017, pp. 9-21.